\RequirePackage[2020-02-02]{latexrelease}
\documentclass[aps,11pt,preprintnumbers,nofootinbib,floatfix]{revtex4}

\usepackage{graphicx}% Include figure files
\usepackage{epsfig}
\usepackage{dcolumn}% Align table columns on decimal point
\usepackage{subfigure}
\usepackage{multirow}
\usepackage{amsmath}

\usepackage{amssymb}
\begin{document}

\title{Compactified extra dimension and entanglement island as clues to quantum gravity}

%%%% To generate auto affiliation numbers please use \author{}\affil{} command
\author{Tran N. Hung}
\email{hung.tranngoc@phenikaa-uni.edu.vn}  
\affiliation{Phenikaa Institute for Advanced Study and Faculty of Fundamental Sciences, Phenikaa University, Yen Nghia, Ha Dong, Hanoi 12116, Vietnam}
\author{Cao H. Nam}
\email{nam.caohoang@phenikaa-uni.edu.vn}  
\affiliation{Phenikaa Institute for Advanced Study and Faculty of Fundamental Sciences, Phenikaa University, Yen Nghia, Ha Dong, Hanoi 12116, Vietnam}
\date{\today}

\begin{abstract}
We show that the compactified extra dimension and the emergence of the island can provide clues about quantum gravity because their combination can solve the deepest puzzles of black hole physics. Suppose that the time dimension and the extra dimension compactified on a circle are symmetric under \emph{double Wick rotation}, the curvature singularity would be removed due to the end of spacetime as a smooth bubble hidden behind the event horizon. The smooth bubble geometries can also be interpreted as microstates leading to the Bekenstein-Hawking entropy because the smooth bubble geometries live in the same region of mass and charge as the black string. In addition, by applying the quantum extremal surface prescription, we show the emergence of the island at late times of the black string evaporation where it is located slightly outside the event horizon. Due to the dominant contribution of the island configuration, the entanglement entropy of the radiation grows no longer linearly in time but it reaches a finite value that is twice the Bekenstein-Hawking entropy at the leading order. This transition shows the information preservation during the black string evaporation. Furthermore, we calculate the Page time which determines the moment of the transition between the linearly growing and constant behaviors of the entanglement entropy as well as the scrambling time corresponding to the information recovery time of the signal falling into the black string.
\end{abstract}

\maketitle

\section{Introduction}
General relativity (GR) predicts the black hole solutions with a curvature singularity surrounded by an event horizon and they have been directly confirmed by the experimental observations \cite{1,2,3}. This singularity is unphysical because the spacetime curvature and the densities become infinite there. Hence, the presence of singularity in the black hole solutions has been perceived as signaling the breakdown of GR in extreme conditions \cite{Hawking1973}. In addition, another important question is about a microscopic description of the black hole entropy: according to statistical mechanics, the black hole entropy would be determined by the number of quantum microstates $\Omega$ as $S_{\text{BH}}=\log\Omega$ which resemble the black hole, but what are the degrees of freedom accounting for the microstates of the black hole entropy? In particular, the black hole would evaporate due to Hawking radiation which is black body radiation \cite{Hawking1975}. This means that if a black hole is formed from the collapse of the matter in a quantum state corresponding to zero entropy, the final state of the black hole evaporation would be in a thermal state corresponding to infinite entropy, leading to the information loss paradox \cite{Hawking1976}. It is widely believed that all issues above can be solved in a consistent theory of quantum gravity that describes how gravity behaves in the short-distance regime where the quantum effects cannot be ignored.

Unfortunately, the obstacles regarding the conceptual and technical aspects have prevented attempts to develop a consistent theory of quantum gravity which is thus lacking. However, it is reasonable to expect that quantum gravity would leave clues in the low-energy regime which provide a bridge from quantum gravity to general relativity or classical gravity. The compactified extra dimensions may be an important clue to seeking a consistent theory of quantum gravity. This is because they are the essential ingredients in constructing superstring/M theory which is regarded as a leading candidate for a quantum theory of gravity. And they may also offer one of the most beautiful and attractive ways towards a geometric unification of gravity with the non-gravitational interactions \cite{Bailin1987,Overduin1997} or dark matter \cite{Nam2022c} as well as provide the phenomenological models for the open problems in astrophysics, cosmology, and particle physics \cite{Antoniadis1990,Dienes1998,Arkani-Hamed1998,Antoniadis1998,Randall1999,Arkani-Hamed2000,Rizzo2018a,Rizzo2018b,Nam2019,Ikeda2019,Cardoso2019,Brax2019,Nam2020a,Sengupta2020,Nam2020b,Ygael2020,Rizzo2021,Nam2021,Nortier2021,Khan2022,Mirzaev2022}. The black hole solutions have been found in the situation that the geometry of the compactified extra dimensions is a $n$-dimensional torus $T^n$ where they possess a horizon with the topology $S^2\times T^n$. In particular, for $n=1$ the black hole solution is called the black string, which is extensively investigated in the literature \cite{Miyamoto2006,Kunz2009,Brihaye2010,RBMann2011,Tannukij2017,Giacomini2018,SHHendi2021,YKLim2021}. Like its counterpart in GR, the black hole solution with the compactified extra dimensions has a curvature singularity (that is different from the point and surrounded by an event horizon) and a temperature associated with Bekenstein-Hawking entropy.

Recently, Bah and Heidmann showed that the compactified extra dimensions could solve some issues of the black hole, thus providing insights into quantum gravity. By considering the vacuum solutions symmetric under double Wick rotation, they found a bubble behind the horizon where the spacetime ends at the bubble's radius \cite{8}. As a result, the usual unphysical singularity does not appear or in other words, the black hole solution, in this case, is regular. In addition, it provides microstate geometries which can be identified as some of the degrees of freedom corresponding to the black hole entropy, which can be coherent enough to be classically described through geometric transitions. Although the investigation of the compactified extra dimensions in Ref. \cite{8} can provide a potential solution for two of three very robust issues of black hole physics, it cannot describe the time evolution of the black hole evaporation compatible with the unitarity principle of quantum mechanics or the Page curve \cite{Page1993a,Page1993b}. This means that the compactified extra dimensions are insufficient to exhibit clues of quantum gravity in the low-energy regime and we need other clues beyond the extra dimensions.

Interestingly, some recent works have shown that the gravitational Euclidean path integral (which is of the main methods to explore quantum gravity) with new saddle points which are the replica wormholes leads to the emergence of the island configuration, which allows for deducing the Page curve \cite{Almheiri2020,Penington2019}. The islands $I$ are some regions that appear in the complementary of the radiation region $R$ assumed to be far away from the black holes. Its boundaries extremize the generalized entropy functional and thus are called the extremal surfaces. In the presence of the island configuration, the entanglement entropy of the Hawking radiation is computed as follows \cite{Ryu2006,Hubeny2007,Engelhardt2015,Faulkner2013,Wall2014,Almheiri2019,Penington2020,Mahajan2019,Akers2020}
\begin{equation}
S(R)=\min \left\lbrace  \exp \left[\dfrac{\mathcal{A}(\partial I)}{4G_{N}}+S_{\text{mat}}(R\cup I) \right] \right\rbrace  ,
\label{eq:ent}
\end{equation}
where $G_N$ is the gravitational constant, $ \partial I $ denotes the island boundaries, $ \mathcal{A}(\partial I) $ refers to the total area of $ \partial I $, and $ S_{\text{mat}} $ is the von Neumann entropy of the quantum fields on the radiation region and the islands.

The island rule has attracted enormous attention in calculating the entanglement entropy of the Hawking radiation and the corresponding Page curve for various black hole geometries. The entanglement entropy is studied via the island rule in the context of Jackiw-Teitelboim (JT) gravity \cite{24} and the different 2D black hole solutions \cite{25,26,JLin2022}. The extensions of the island proposal for the higher-dimensional black holes using the approximation of the 2D conformal field theory (CFT) have been studied concerning the Schwarzschild black holes \cite{27, Ageev}, the Reissner-Nordstr\"{o}m (RN) black holes \cite{LiWang2021,KimNam2021}, the charged/neutral dilaton black holes \cite{Karananas2021,YuGe2021,Ahn2021}, the Kaluza-Klein black holes \cite{13}, nonextremal asymptotically flat or AdS black hole \cite{SHe2022}, and the rotating black holes \cite{35,36}. Many interesting works have indicated the entanglement island in moving mirrors models \cite{Kumar2022}, the higher-dimensional black hole in the presence of an end-of-the-world brane defining the time-dependent region of the effective Hawking radiation \cite{CChou2022}, the braneworld geometries \cite{Krishnan2021,Afrasiar2022,Afrasiar2023,Perez-Pardavila2023}, flat-space cosmologies \cite{Azarnia2021}, de Sitter spacetime \cite{Balasubramanian2021,Azarnia2022,Goswami2022,Sybesma2022}, and the context of the AdS/BCFT correspondence \cite{Karch2022,DBasu2022,Randall2022}. Applying the extremal surface technique, it was pointed that the island configuration also emerges to make the entanglement entropy remain constant in the late times in the gravity theories with the massive graviton \cite{Geng2020,Geng2021,Geng2022,HNam2022}, the holographic axion gravity \cite{Hosseini2022}, the gravity including the higher derivative terms \cite{31,32,33}, and deformed JT gravity \cite{CYLu2022}.

In this work, we will apply the quantum extremal surface technique to investigate the entanglement entropy of the Hawking radiation and the corresponding Page curve for the regular black solution found in Ref. \cite{8} with a bubble behind the event horizon. The calculation finds that the entanglement island emerges at late times and its radius rises with the size of the bubble. The emergence of the island configuration prevents the entanglement entropy of the Hawking radiation from growing infinitely but instead it is nearly a constant value which is twice the Bekenstein-Hawking entropy after the Page time. In this way, the time evolution of the entanglement entropy of the Hawking radiation follows the Page curve. As a result, the evaporation process of the black hole respects the unitarity principle. This clearly implies that the compactified extra dimensions and the entanglement island can together lead to the solutions for three important issues of black hole physics (namely the unphysical curvature singularity, microstates of the Bekenstein-Hawking entropy, the unitary time evolution of the black hole evaporation) in some sense. Therefore, they provide essential clues to constructing a consistent theory of quantum gravity.

This paper is organized as follows. In Sec. \ref{sec:2}, we briefly review the regular black string solutions found in Ref. \cite{8}. In Sec. \ref{sec:3}, we calculate the entanglement entropy of Hawking radiation for the five-dimensional black string in the configurations without and with the island. We also determine the Page and scrambling times, thereby recovering the Page curve of the entanglement entropy. In addition, we investigate the entanglement entropy for the higher dimensional black strings and evaluate the effects of the size of the bubble on the radius of the island. In Sec. \ref{CHD}, we apply the island formula to calculate the entanglement entropy in the case of higher-dimensional black strings with a compactified extra dimension. In the last section, we conclude our main results.

\section{Black string solution without curvature singularity}
\label{sec:2}

In this section, we briefly present the black string solution without curvature singularity due to the presence of a smooth bubble behind the event horizon \cite{8}. The spacetime ends at the radius of the bubble and as a result, the region of the usual curvature singularity is naturally removed. 

We consider the Einstein-Maxwell system in five dimensions whose action is given as follows
\begin{equation}
S=\int d^5x\sqrt{-g}\left(\frac{1}{2\kappa_{5}^{2}}R-\frac{1}{4}F_{\mu\nu}F^{\mu\nu}\right),
\end{equation}
where $ \kappa_{5} $ is the five-dimensional gravitational coupling. We find the spherically symmetric solution of the system with a magnetic charge, which is described by the following ansatz
\begin{equation}
ds^{2}=-f_{S}(r)dt^{2}+f_{B}(r)dy^{2}+\dfrac{dr^{2}}{h(r)}+r^{2}\left(d\theta^{2}+\sin^{2}\theta d\phi^{2}\right),
\label{eq:metric}
\end{equation}
\begin{equation}
F=P\sin \theta d\theta\wedge d\phi.\label{Maxan}
\end{equation}
Because the fifth dimension is compactified on a circle $S^1$ with the radius $R_y$, the coordinate $y$ is periodic with the period $2\pi R_y$. For the vacuum solutions corresponding to the magnetic flux turned off or $P=0$, one can find two solutions as
\begin{itemize}
\item[(i)] The product of 4D Schwarzchild black hole with $S^1$:
\begin{eqnarray}
\label{eqn:sol1}
f_{S}(r)=h(r)=1-\dfrac{r_{S}}{r},\ \ \quad f_{B}(r)=1.
\end{eqnarray}
This solution has a timelike Killing vector $ \partial_{t} $ which shrinks at the event horizon $r=r_S$.
\item[(ii)] The smooth geometry solution corresponding to a static bubble of nothing at $r=r_B$:
\begin{eqnarray}
\label{eqn:sol2}
f_{B}(r)=h(r)=1-\dfrac{r_{B}}{r},\ \ \quad f_{S}(r)=1.
\end{eqnarray}
This solution has a spacelike Killing vector $ \partial_{y} $ which shrinks at $r=r_B$.
\end{itemize}
We suppose that these solutions are both vacuum solutions, which means that they are symmetric under the double Wick rotation $ (t, y, r_{S}, r_{B})\rightarrow (iy, it, r_{B}, r_{S}) $. This can be solved by turning on the magnetic fluxes, which leads to
\begin{eqnarray}
    \quad h(r)&=&f_{B}(r)f_{S}(r),\label{hr-funct}\\
P&=&\pm\dfrac{1}{\kappa_{5}}\sqrt{\dfrac{3r_{S}r_{B}}{2}}.\label{mag-char}
\end{eqnarray}
where $f_S(r)=1-r_S/r$ and $f_B(r)=1-r_B/r$. There are two coordinate singularities appearing at $ r=r_{S} $ and $ r=r_{B} $. 

The solution that is given by Eqs. (\ref{eq:metric}) and (\ref{Maxan}) with $h(r)$ and $P$ given by Eqs. (\ref{hr-funct}) and (\ref{mag-char}), respectively, can lead to two types of topology depending on the values of $r_S$ and $r_B$. The solution is either a massive magnetic bubble for $ r_{B}>r_{S} $ or a black string of the magnetic charge for $ r_{S}\geq r_{B} $. 

\textbf{The smooth bubble solution (topological star).}  In the case of $ r_{B}>r_{S} $, the compactified extra dimension shrinks to a zero size at $r=r_B$, which implies the end of the spacetime there. The spacetime geometry near $r_B$ is
\begin{eqnarray}
    ds^2=-\frac{r_B-r_S}{r_B}dt^2+r^2_B\left[d\rho^2+\frac{r_B-r_S}{4r^3_B}\rho^2dy^2+d\theta^{2}+\sin^{2}\theta d\phi^{2}\right],
\end{eqnarray}
where $\rho\equiv2\left[(r-r_B)/(r_B-r_S)\right]^{1/2}\rightarrow0$. In general, the local metric corresponding to the $(\rho,y)$ subspace has a conical defect (which describes a localized object within string theory) associated with a topology $ \mathbb{R}^{2}/\mathbb{Z}_{k}$ with $k\in \mathbb{Z}_{+}$. Then, one can determine the product of the quantum number $k$  and the radius of the compactified extra dimension in terms of the parameters of the bubble solution as follows
\begin{eqnarray}
k^2R^2_y=\frac{4r^3_B}{r_B-r_S}.
\end{eqnarray}
The mass and charge of the topological star are
\begin{eqnarray}
M=\dfrac{2\pi r_{B}}{\kappa_{4}^{2}}\left( 3-8\dfrac{r_{B}^{2}}{k^{2}R_{y}^{2}} \right),\quad Q_{m}^{2}=\dfrac{3r_{B}^{2}}{2\kappa_{4}^{2}}\left( 1-4\dfrac{r_{B}^{2}}{k^{2}R_{y}^{2}} \right).  
\end{eqnarray}
We have an upper bound of the radius of the bubble as $ 2r_{B}\leq kR_{y} $. The charge vanishes at $ r_{B}=kR_{y}/2 $ and the solution is a vacuum bubble of nothing.

\begin{figure}[t]
  \includegraphics[scale=0.3]{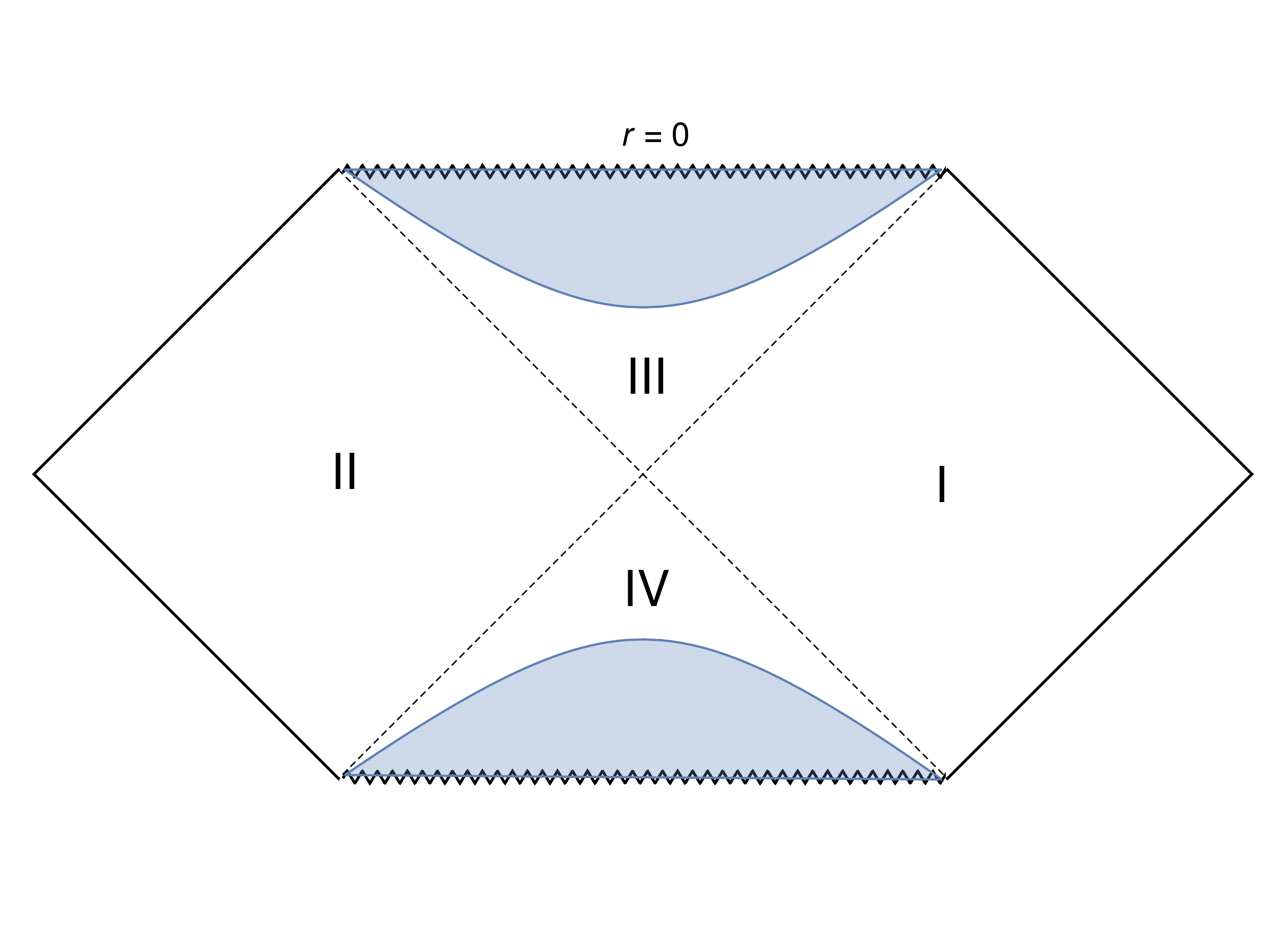}
  \caption{Penrose diagram of the regular black string. The gray regions refer to the ends of spacetime as a smooth bubble.}
  \label{fig:f1}
\end{figure}

\textbf{The black string.}  In the case of $ r_{S}>r_{B} $, there are two coordinate singularities at $ r=r_{S} $ and $ r=r_{B} $. The horizon appears at the first singularity. In order to see the topology of the horizon, we consider the near-horizon geometry as follows
\begin{eqnarray}
ds^2=-\frac{r_S-r_B}{4r^3_S}\rho^2dt^2+d\rho^2+\frac{r_S-r_B}{r_S}dy^2+r^2_S\left(d\theta^{2}+\sin^{2}\theta d\phi^{2}\right),
\end{eqnarray}
where $\rho\equiv2\left[(r-r_S)/(r_S-r_B)\right]^{1/2}r_S\rightarrow0$. From this local geometry, we find that the horizon of the black string has topology $ S^{1}\times S^{2} $ where the radii for the $ S^1 $ and $S^2$ are $\sqrt{1-r_S/r_B}R_y$ and $r_S$, respectively. The second singularity at $ r=r_{B} $ that is located behind the horizon defines a bubble $S^2$  which is a timelike surface where the local geometry is given by
\begin{eqnarray}
 ds^2=\frac{r_S-r_B}{r_B}dt^2+r^2_B\left[-d\rho^2+\frac{r_S-r_B}{4r^3_B}\rho^2dy^2+d\Omega^2_2\right],
\end{eqnarray}
where $\rho\equiv2\left[(r-r_B)/(r_S-r_B)\right]^{1/2}\rightarrow0$. With the transformation $T=-\rho\cosh(\gamma\varphi)$ and $R=-\rho\sinh(\gamma\varphi)$ where the $2\pi$-periodic angle $\varphi$ is defined as $\varphi=y/R_y$ and the parameter $\gamma$ is given by $ \gamma^{2}=(r_{S}-r_{B})R_{y}^{2}/(4r_{B}^{3}) $, we can be written the two-dimensional line element of the $(\rho,y)$ subspace in terms of the new coordinates $(T,R)$ as follows
\begin{eqnarray}
-d\rho^2+\frac{r_S-r_B}{4r^3_B}\rho^2dy^2=-dT^2+dR^2.
\end{eqnarray}
This $(T,R)$ subspace is a cone in the two-dimensional Minkowski space $\mathbb{R}^{1,1}$. The bubble $S^2$ sits at the apex of the cone where the spacelike Killing vector $\partial_y$ shrinks. The causal structure of the black string is described by the Penrose diagram depicted in Fig. \ref{fig:f1}.
In four-dimensional spacetime, a black string displays a magnetic black hole of mass and charge. This black hole has a horizon at $ r=r_{S} $ corresponding to the Bekenstein-Hawking entropy and the Hawking temperature as follows
\begin{eqnarray}
S_{\text{BH}}&=&\frac{8\pi^{2}}{\kappa_{4}^{2}}\sqrt{r_{S}^{3}(r_{S}-r_{B})},\\
T_\text{H}&=&\frac{1}{4\pi r_S}\sqrt{1-\frac{r_B}{r_S}}.
\end{eqnarray}

By studying the phase space in the plane of the four-dimensional mass and charge, it was pointed out that there is a regime of the mass and charge where the smooth bubble solution, black string, and RN black hole can live together \cite{8,BahHeidmann2021}. On the other hand, the mass and the charge of the smooth bubble solutions are the same as those of the black string and the four-dimensional charged black holes. Hence, it is interesting that the smooth bubble geometries can be realized as microstates that lead to black hole entropy.

\section{Entanglement entropy}
\label{sec:3}

In this section, we shall evaluate the entanglement entropy of the Hawking radiation emitted from the regular black string discussed above in the s-wave approximation. We consider the contribution of the configurations without and with the islands to the entanglement entropy. It is easy to show the infinity of the entanglement entropy of the radiation in the configuration without any island. Otherwise, we will indicate that the entanglement entropy is finite at the late times in the configuration of a single island. Consequently, the Page curve describing the evaporation of the black hole is consistent with the unitarity principle.

In order to calculate the entanglement entropy, first we need to introduce the tortoise coordinate as follows
\begin{equation}
\begin{split}
r_{*} (r)&= \int \dfrac{dr}{\sqrt{f_{B}(r)}f_{S}(r)}\\
&=r\sqrt{1-\dfrac{r_{B}}{r}}+\left(\dfrac{r_{B}}{2}+r_{S} \right)\log \left[ -1+\dfrac{2r}{r_{B}}\left(1+\sqrt{1-\dfrac{r_{B}}{r}} \right) \right]\\
&\quad -\dfrac{r_{S}^{3/2}}{\sqrt{r_{S}-r_{B}}}\log\left[\dfrac{2r}{r-r_{S}}\sqrt{1-\dfrac{r_{B}}{r}}+\dfrac{r(r_{B}-2r_{S})+r_{B}r_{S}}{(r_{S}-r)\sqrt{r_{S}(r_{S}-r_{B})}} \right].
\end{split}
\end{equation}
Then, we define the Kruskal coordinate as
\begin{equation}
U \equiv -e^{-\kappa (t-r_{*})}, V \equiv e^{\kappa (t+r_{*})},
\end{equation}
where $ \kappa $ is the surface gravity of the black string and it is defined by
\begin{equation}
\kappa=\dfrac{1}{2r_{S}}\sqrt{1-\dfrac{r_{B}}{r_{S}}}.
\end{equation}
With these coordinate transformations, the line element (\ref{eq:metric}) is rewritten in terms of the Kruskal coordinate as
\begin{equation}
ds^{2}=-W^{2}(r)dUdV+f_{B}(r)dy^{2}+r^{2}d\Omega^{2},
\end{equation}
where the conformal factor reads
\begin{equation}
W^{2}(r)=\dfrac{f_{S}(r)}{\kappa^{2}e^{2\kappa r_{*}}}.
\end{equation}

\label{sub:noisl}
\begin{figure}[b]
  \includegraphics[scale=0.3]{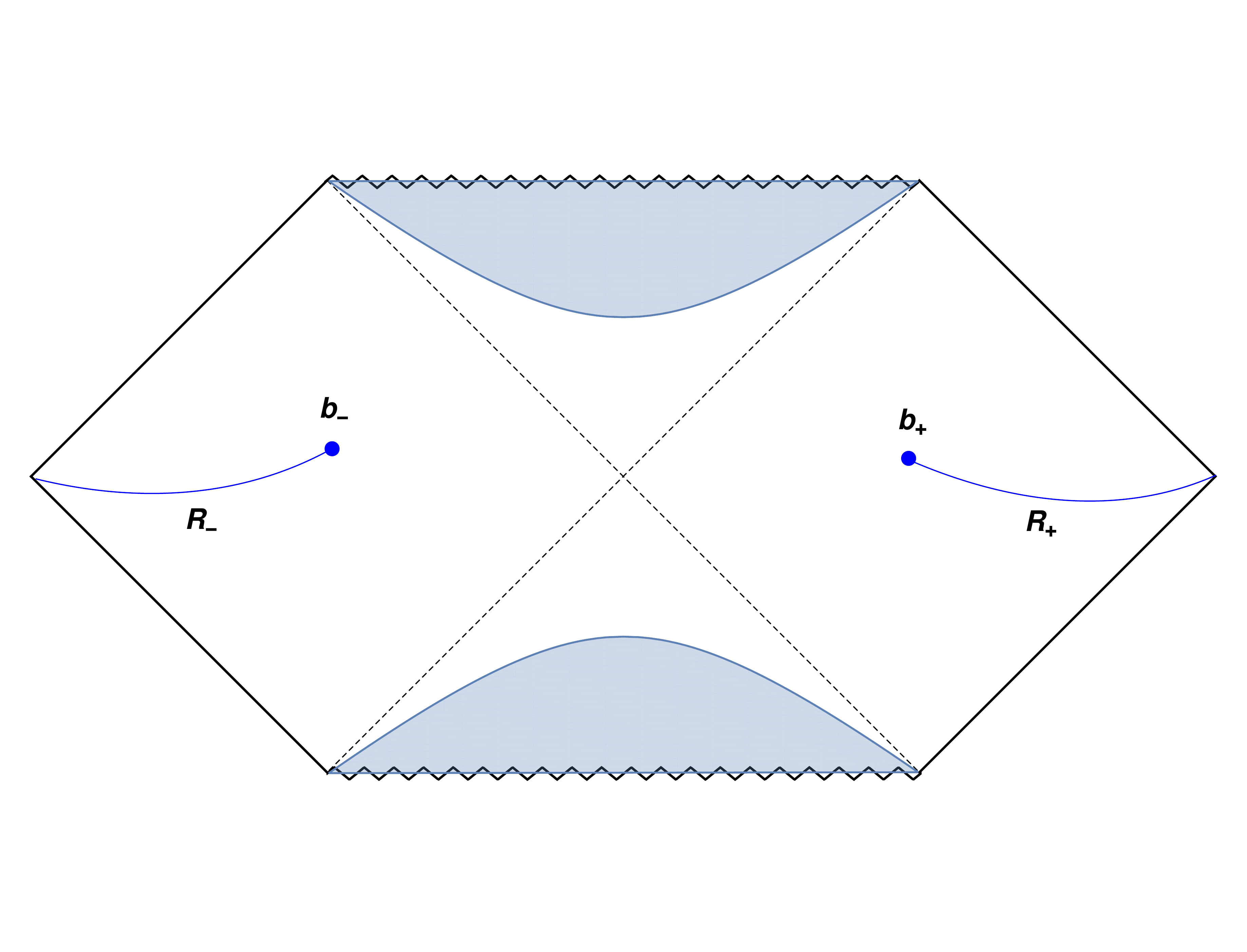}
  \caption{The Penrose diagram of the eternal non-extremal regular black string without islands. The radiation region is the union of two parts $R_{\pm}$ whose boundaries are denoted by $b_{\pm}$.}
  \label{fig:f2}
\end{figure}

\subsection{Without islands}

Let us now calculate the entanglement entropy of the radiation in the case of the configuration without islands. The Penrose diagram of the regular black string without the islands is shown in Fig. \ref{fig:f2}. The radiation region includes two regions $ R_{-} $ and $ R_{+} $, which are located in the left and right wedges, respectively. The cutoff surfaces of the regions $ R_{-} $ and $ R_{+} $ are denoted by $ b_{-} $ and $ b_{+} $, respectively. These endpoints $ b_{\pm} $ have the coordinates $ (t_{b},b) $ for $ b_{+} $ and $ (-t_{b}+i\beta/2,b) $ for $ b_{-} $, where $ \beta $ is the inverse of the Hawking temperature. Assuming that the distance between two endpoints $ b_{-} $ and $ b_{+} $ is large enough compared to the size of these boundaries, we can use the two-dimensional conformal field theory (CFT) to approximately calculate the entanglement entropy. For the situation where the initial state of the whole system is in the pure state, the entanglement entropy of the radiation region outside $ [b_{-},b_{+}] $ is equal to the one within the interval. In this way, the entanglement entropy is calculated as follows \cite{Fiola1994,Calabrese2004}
\begin{equation}
\begin{split}
S_{R} &=\dfrac{c}{3}\log d(b_{+},b_{-})\\
&=\dfrac{c}{6}\log\left[W(b_{+})W(b_{-})(U(b_{-})-U(b_{+}))(V(b_{+})-V(b_{-})) \right] \\
&=\dfrac{c}{6}\log \left[4W^{2}(b)e^{2\kappa r_{*}(b)}\cosh^{2}(\kappa t_{b}) \right] \\
&\simeq \dfrac{c}{6} \log\left[ (1-\dfrac{r_{S}}{b})\cosh^{2}(\kappa t_{b})\right] ,\\
\end{split}
\end{equation}
where $ c $ is the central charge of the two-dimensional CFT and $ d(b_{+},b_{-}) $ is the distance between $ b_{+} $ and $ b_{-} $.

At early times, we use the approximation $ t_{b}\ll 1/\kappa $ to deduce
\begin{equation}
S_{R}\simeq\dfrac{c}{6}\log(1-\dfrac{r_{S}}{b})+\dfrac{c}{6}(\kappa t_{b})^{2}.
\end{equation}
In the above formula, the first term is the initial entanglement entropy of the radiation region which is independent of the radius of the bubble and the second term shows the quadratic evolution of the entanglement entropy with time. For late times corresponding to the approximation $ t_{b}\gg 1/\kappa $, we find the corresponding entanglement entropy as
\begin{equation}
S_{R}\simeq \dfrac{c}{3}\kappa t_{b}.
\label{eq:noisl}
\end{equation}
This result exhibits that the entanglement entropy of the Hawking radiation grows linearly in time and thus it will reach the infinite value as $ t_{b}\rightarrow \infty $. Clearly, this result conflicts with the finiteness of the total von Neumann entropy of the black string. However, this conflict will be resolved when taking into account the island configuration evaluated in the next section, where it can reproduce the correct Page curve.
\subsection{With an island}
\begin{figure}[b]
  \includegraphics[scale=0.4]{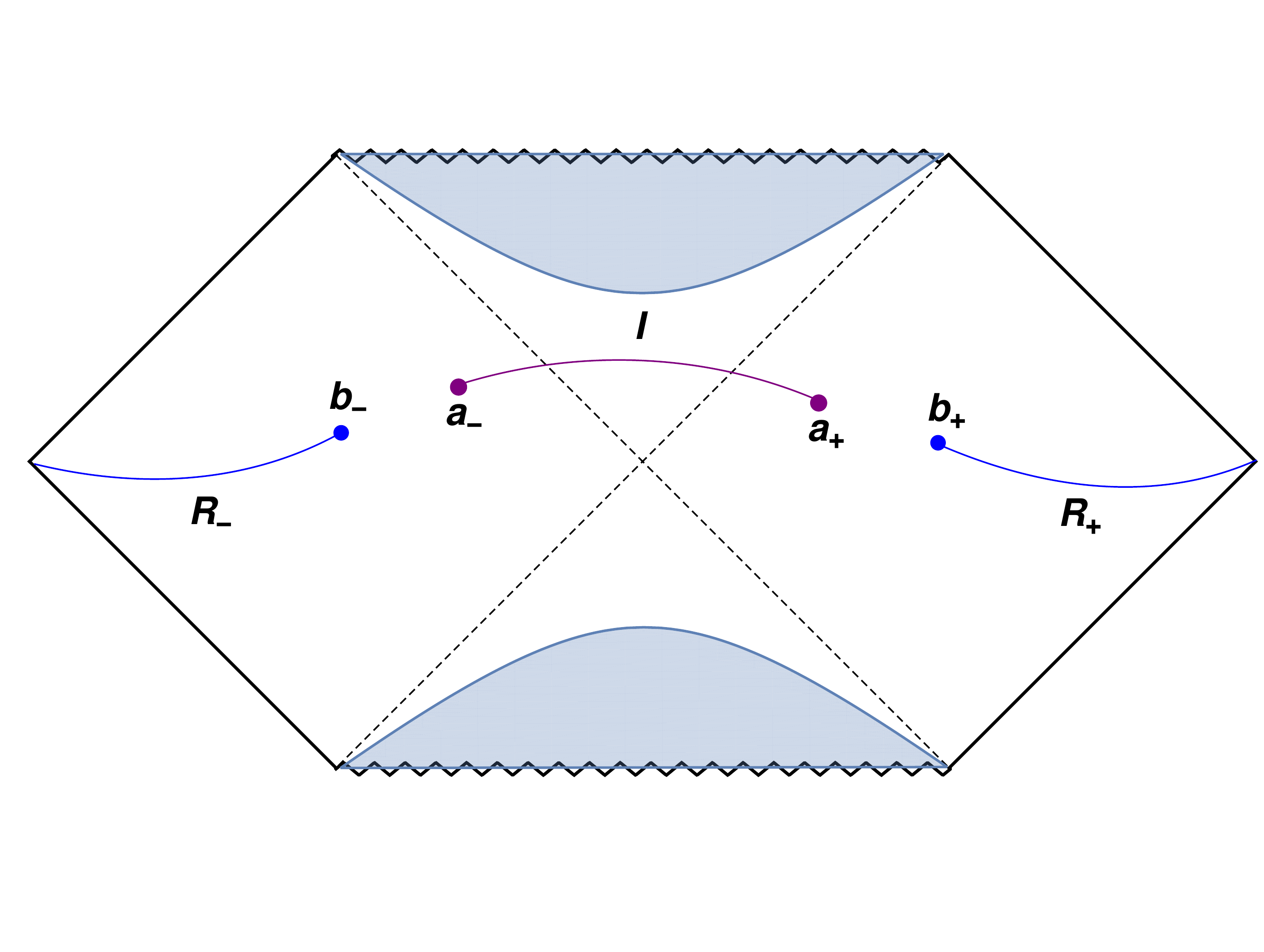}
  \caption{The Penrose diagram of the eternal non-extremal regular black string with an island. $ I $ refers to the island which has two boundaries denoted by $ a_{\pm} $.}
  \label{fig:f3}
\end{figure}

When including the island configuration, the generalized entropy is a sum of two contributions where the first contribution is relevant to the gravitational part that is proportional to the total area of the island boundaries and the second contribution is the entanglement entropy of the matter. It is given as
\begin{equation}
S_{\text{gen}} = \dfrac{16\pi^{2}}{\kappa_{4}^{2}}\sqrt{a^{3}(a-r_{B})}+S_{\text{mat}}(R_{-}\cup R_{+}\cup I),
\end{equation}
where the first term is the Bekenstein-Hawking entropy of the island and the second term is the entanglement entropy of the matter (which lives on the radiation region and the island) is calculated by the following formula 
\begin{equation}
\begin{split}
S_{\text{mat}}&=\dfrac{c}{3}\log \left[ d(a_{+},a_{-})d(b_{+},b_{-})\right]+\dfrac{c}{3}\log \left[\dfrac{d(a_{+},b_{+})d(a_{-},b_{-})}{d(a_{+},b_{-})d(a_{-},b_{+})}\right] \\
&=\dfrac{c}{6}\log\left[ 2^{4}W^{2}(a)W^{2}(b)e^{2\kappa (r_{*}(a)+r_{*}(b))}\cosh^{2}(\kappa t_{a})\cosh^{2}(\kappa t_{b})\right] \\
&\quad+\dfrac{c}{3}\log\left[\dfrac{\cosh(\kappa (r_{*}(a)-r_{*}(b)))-\cosh(\kappa (t_{a}-t_{b}))}{\cosh(\kappa (r_{*}(a)-r_{*}(b)))+\cosh(\kappa (t_{a}+t_{b}))} \right].
\end{split}
\end{equation}
By extremizing the generalized entropy with respect to the temporal and spatial location of the island boundaries, we find the corresponding minimum value. If it exists, this value would be identified as the entanglement entropy of the Hawking radiation. In the following, we study the behavior of the entanglement entropy at early and late times of the evaporation process of the black string.

\textbf{Early times.}

At early times, the entanglement entropy of Hawking radiation is small. Accordingly, we expect the island to lie deep inside the black string. In the approximation that the cutoff surface is far away from the event horizon, given by
\begin{equation}
r_{S}\ll b;\quad t_{a},t_{b}\ll 1/\kappa \ll r_{*}(b)-r_{*}(a),
\end{equation}
we can obtain the generalized entropy as
\begin{equation}
\begin{split}
S_{\text{gen}}&\simeq \dfrac{16\pi^{2}}{\kappa_{4}^{2}}\sqrt{a^{3}(a-r_{B})}+\dfrac{c}{6}\log\left[ |(f_{S}(a))|\cosh^{2}(\kappa t_{a})\right] + \cdots\\
&\simeq \dfrac{16\pi^{2}}{\kappa_{4}^{2}}\sqrt{a^{3}(a-r_{B})}+\dfrac{c}{6}\log(\dfrac{r_{S}}{a}-1)+\dfrac{c}{6}(\kappa t_{a})^{2}+\cdots.
\end{split}
\end{equation}
The terms that are relevant to $ t_{b}$ and $r_{*}(b)$ only are ignored without affecting the extremizing condition of the generalized entropy. The partial derivative of the generalized entropy with respect to the position of the island boundary is given as follows
\begin{equation}
\begin{split}
\dfrac{\partial S_{\text{gen}}}{\partial a}&=\dfrac{16\pi^{2}}{\kappa_{4}^{2}}\dfrac{a(4a-3r_{B})}{2\sqrt{a(a-r_{B})}}-\dfrac{c}{6}\dfrac{r_{S}}{a(r_{S}-a)}\\
&\simeq \dfrac{16\pi^{2}}{\kappa_{4}^{2}}\dfrac{a(4a-3r_{B})}{2\sqrt{a(a-r_{B})}}-\dfrac{c}{6}\dfrac{1}{a},
\end{split}
\end{equation}
where we use the approximation $ a\ll r_{S} $ in the second line. From the extremizing condition $ \dfrac{\partial S_{\text{gen}}}{\partial a}=0 $ we can deduce the following relation
\begin{equation}
a \sim \sqrt{c\kappa_{4}^{2}}\sim l_{P},
\label{eq:isle}
\end{equation}
where $ l_{P} $ is the Planck length. The expression (\ref{eq:isle}) implies the presence of an island of the Planck scale inside the black string. However, we observe that the bubble radius must be much larger than the Planck length because the bubble geometries which are expected to identify as the microstates are not quantumly but the classical description coming from the decoherence of some quantum microstates. In this sense, the island which appears at early times should be located in the gray regions (given in Fig. \ref{fig:f3}), which correspond to the end of spacetime at the bubble. In addition, the Planck size of the Planck also conflicts with the calculating approach in the derivation of the island rule, which requires the upper cutoff length scale much larger than the Planck length. Therefore, these facts indicate that no island actually emerges at early times. Consequently, the entanglement entropy would be determined by the geometry configuration without the island.

\textbf{Late times.}

At the late stage of the evaporation process of the black string, the radiation entering the cutoff surface becomes more and more. The contribution of the radiation thus grows with time. We should expect that the coarse-grain entropy increases linearly, but the fine-grain entropy will reach a finite value to satisfy the unitarity of quantum theory. 

At late times, we can assume that the radiation region is far from the horizon. Therefore, we use the following approximation
\begin{equation}
1/\kappa \ll r_{*}(b)-r_{*}(a)\ll t_{a},t_{b},\label{latet-appr}
\end{equation}
which leads to 
\begin{equation}
\begin{split}
\cosh \kappa t_{a,b}\simeq \dfrac{1}{2}e^{\kappa t_{a,b}}, \\
\cosh \kappa(t_{a}+t_{b})\gg \cosh \kappa(r_{*}(b)-r_{*}(a)).
\end{split}
\end{equation}
With this approximation, we can calculate the time-dependent component of the generalized entropy as
\begin{equation}
\begin{split}
S_{\text{time}}&=\dfrac{c}{3}\log\left[\cosh \kappa t_{a} \cosh \kappa t_{b} .\dfrac{\cosh(\kappa (r_{*}(a)-r_{*}(b)))-\cosh(\kappa (t_{a}-t_{b}))}{\cosh(\kappa (r_{*}(a)-r_{*}(b)))+\cosh(\kappa (t_{a}+t_{b}))}\right] \\
&\simeq \dfrac{c}{3}\log\left[\cosh \kappa (r_{*}(a)-r_{*}(b))-\cosh \kappa (t_{a}-t_{b}) \right] .
\end{split}
\end{equation}
We find that extremizing the generalized entropy with respect to $ t_{a} $ leads to $ t_{a}=t_{b} $. Substituting this result into the generalized entropy, we obtain the first-order approximate formula as
\begin{equation}
\begin{split}
S_{\text{gen}}&= \dfrac{16\pi^{2}}{\kappa_{4}^{2}}\sqrt{a^{3}(a-r_{B})}+\dfrac{c}{6}\log\left[ W^{2}(a)W^{2}(b)\right] +\dfrac{c\kappa}{3}\left(r_{*}(a)+r_{*}(b) \right) \\
&\quad+\dfrac{c}{6}\log\left[ \cosh^{2}(\kappa t_{a})\cosh^{2}(\kappa t_{b})\right]+\dfrac{c}{3} \log\left[\dfrac{\cosh(\kappa (r_{*}(a)-r_{*}(b)))-\cosh(\kappa (t_{a}-t_{b}))}{\cosh(\kappa (r_{*}(a)-r_{*}(b)))+\cosh(\kappa (t_{a}+t_{b}))} \right]\\
&\simeq \dfrac{16\pi^{2}}{\kappa_{4}^{2}}\sqrt{a^{3}(a-r_{B})}+\dfrac{c}{6}\log\left[ W^{2}(a)W^{2}(b)\right]+\dfrac{2c}{3}\kappa r_{*}(b)+ \dfrac{c}{3}\log\left[ 1-2e^{-\kappa(r_{*}(b)-r_{*}(a))} \right] +\cdots\\
&\simeq \dfrac{16\pi^{2}}{\kappa_{4}^{2}}\sqrt{a^{3}(a-r_{B})}+\dfrac{c}{6}\log\left[ f_{S}(a)f_{S}(b)\right]-\dfrac{c}{3}\kappa r_{*}(a)+\dfrac{c}{3}\kappa r_{*}(b)-\dfrac{2c}{3}e^{-\kappa(r_{*}(b)-r_{*}(a))}+\cdots.
\end{split}
\label{eq:gentro}
\end{equation}
This expression of $S_{\text{gen}}$ weakly depends on time, showing that the generalized entropy would have a finite value at late times. This shall induce the convergence of the entanglement entropy at late times instead of growing linearly with time.

When considering the island located near the event horizon, using the extremizing condition $ \dfrac{\partial S_{\text{gen}}}{\partial a}=0 $ we find  
\begin{equation}
a\simeq r_{S}\left[1 + \left(\dfrac{c\kappa_{4}^{2}}{r_{S}^{2}}\right) ^{2}\dfrac{K^{2}}{576\pi^{4}}\dfrac{\sqrt{1-r_{B}/r_S}}{(4-3r_{B}/r_S)^{2}}\right],
\label{eq:israd}
\end{equation}
where $ K $ is defined as follows
\begin{equation}
K\equiv\exp\left\{ \kappa\left[\sqrt{r_{S}(r_{S}-r_{B})}+\left( r_{S}+\frac{r_{B}}{2}\right)\log\left[-1+\dfrac{2r_{S}}{r_{B}}\left( 1+\sqrt{1-\dfrac{r_{B}}{r_{S}}}\right)  \right]-r_{*}(b)\right] \right\} .
\end{equation}
Because of the extremely small factor $ c\kappa_{4}^{2}/r_{S}^{2} $, the second term in (\ref{eq:israd}) is very small but positive. This implies that the location of the island is nearly outside the event horizon of the black string. In addition, the effect of the bubble on the island location predominantly comes from the term $\sqrt{1-r_{B}/r_S}(4-3r_{B}/r_S)^{-2}$ due to $\ln K$ suppressed strongly by the Planck length.
\begin{figure}[htp]
  \includegraphics[scale=0.5]{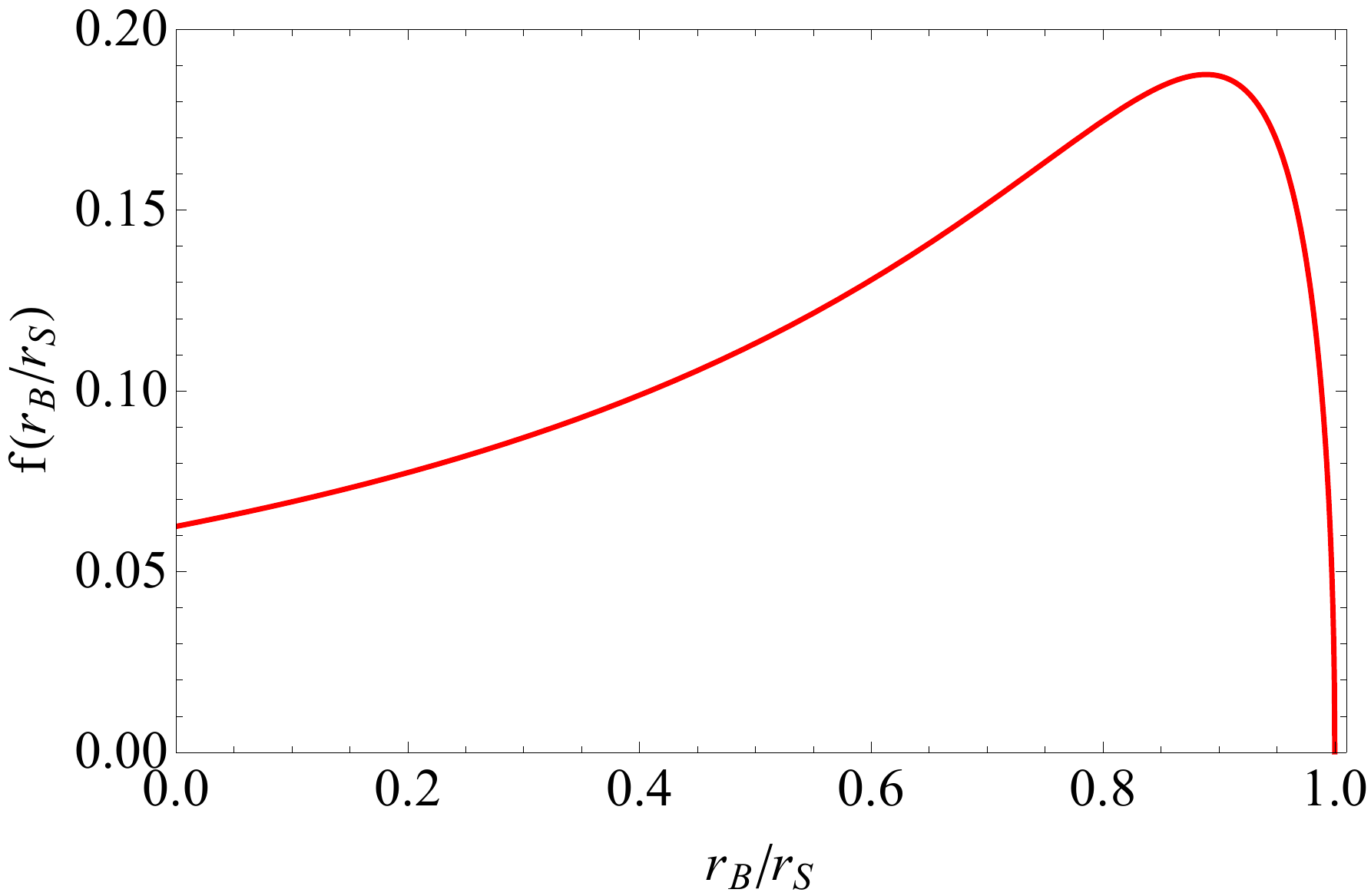}
  \caption{The behavior of the function $f(r_B/r_S)\equiv\sqrt{1-r_{B}/r_S}(4-3r_{B}/r_S)^{-2}$ in terms of $r_B/r_S$.}
  \label{f-rB-rS}
\end{figure}
From the behavior of $\sqrt{1-r_{B}/r_S}(4-3r_{B}/r_S)^{-2}$ in terms of $r_B/r_S$ as depicted in Fig. \ref{f-rB-rS}, we observe that for $r_B/r_S<8/9$ increasing the ratio of the bubble radius to the event horizon would make the island shifted outside the black string. On the contrary, in the regime of $r_B/r_S>8/9$, the growth of this ratio would lead to the island being closer to the black string.

Finally, by substituting the location of the island just obtained above into the approximate expression of the generalized entropy (\ref{eq:gentro}), we derive the entanglement entropy as follows
\begin{eqnarray}
S_{\text{EE}}&\simeq&\frac{16\pi^{2}}{\kappa_{4}^{2}}\sqrt{r_{S}^{3}(r_{S}-r_{B})}+\frac{c\kappa}{3}\left[\sqrt{b(b-r_{B})}-\sqrt{r_{S}(r_{S}-r_{B})}\right]\nonumber \\
&&+\dfrac{c\kappa}{3} \left(r_{S}+\dfrac{r_{B}}{2} \right)\log\dfrac{2(b+\sqrt{b(b-r_{B})})-r_{B}}{2(r_{S}+\sqrt{r_{S}(r_{S}-r_{B})}-r_{B})}  + \mathcal{O}\left(\dfrac{c\kappa_{4}^{2}}{r_{S}^{2}} \right) .
\label{eq:yisl}
\end{eqnarray}
The first term in the above expression is twice the Bekenstein-Hawking entropy of the black string, which comes from the contribution of the island configuration. The second and third terms come from the quantum nature of the matter fields, which are suppressed by the inverse Planck scale $\kappa$. Hence, the terms which are proportional to the nonzero powers of $\kappa$ are very small compared to the first term and thus they are negligible. As a result, the entanglement entropy is approximately given as $S_{\text{EE}}\simeq16\pi^{2}\sqrt{r_{S}^{3}(r_{S}-r_{B})}/\kappa_{4}^{2}$ which is twice the Bekenstein-Hawking entropy of the black string and constant in time. Therefore, the configuration with an island can resolve the information conservation issue during the evaporation of the black strings.

\subsection{Page time and scrambling time}
\begin{figure}[b]
  \includegraphics[scale=0.4]{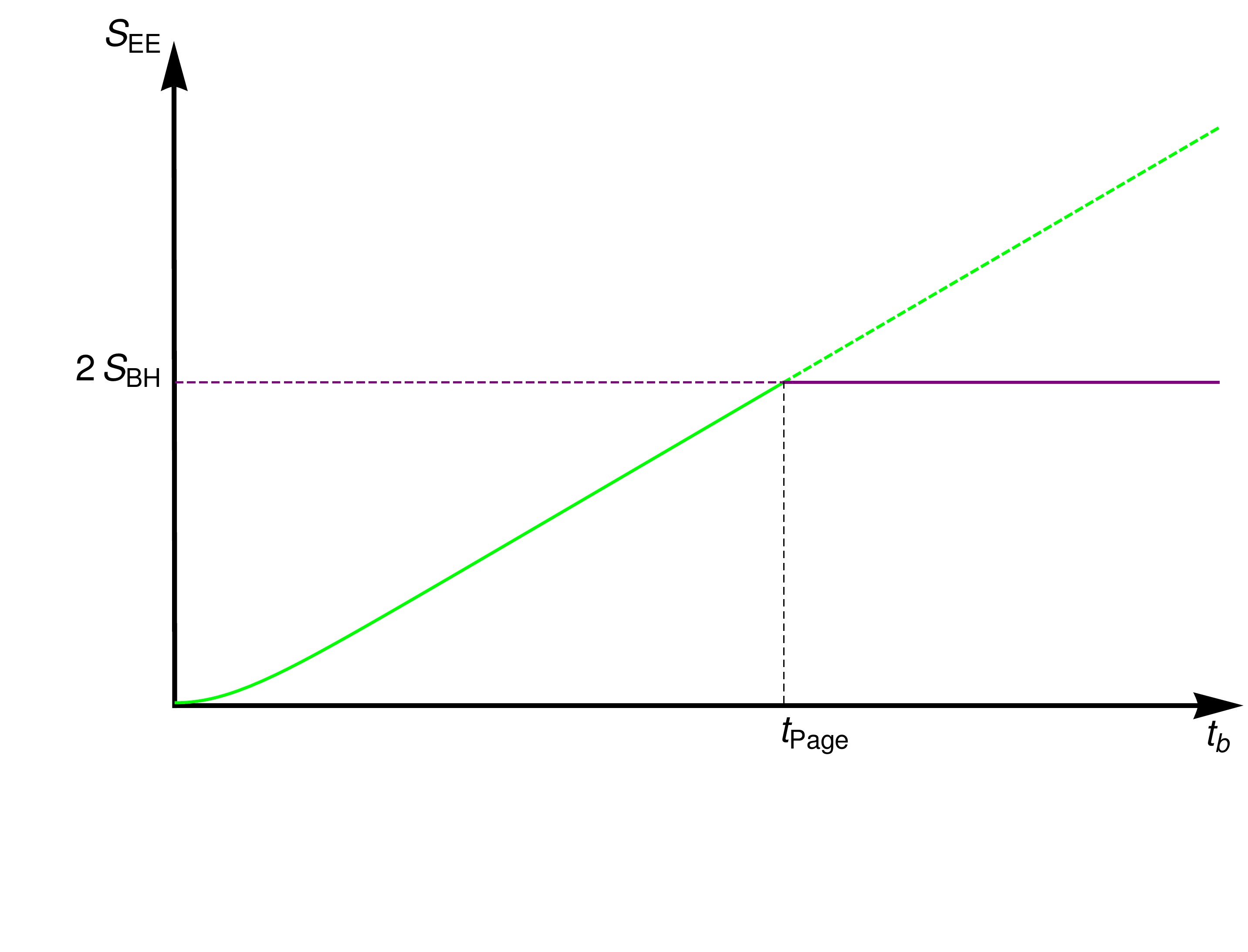}
  \caption{The Page curve for the entanglement entropy of the five-dimensional black string. The solid and dashed green lines represent the time evolution of the entanglement entropy for the configuration without an island. The solid purple line refers to the entanglement entropy at the late times for the configuration with an island.}
  \label{fig:f4}
\end{figure}

The above calculations suggest the time evolution of the entanglement entropy as shown in Fig. \ref{fig:f4}. When the configuration without the island is applied at the early times of the black string evaporation, the entanglement entropy increases linearly with time. However, with the appearance of the island at late times, the entanglement entropy approaches a maximum value at the moment called the Page time and does not change in time. This corresponds to the dominance of the configuration with the island, which is formed near the black hole horizon. From Fig. \ref{fig:f4}, we can approximately determine the Page time at the intersection point between the purple line and the green curve, which corresponds to the configurations without and with the island, respectively. Accordingly, by using Eqs. (\ref{eq:noisl}) and (\ref{eq:yisl}) we obtain the following equation
\begin{equation}
\dfrac{c\kappa}{3}t_{\text{Page}}\simeq\dfrac{16\pi^{2}}{\kappa_{4}^{2}}\sqrt{r_{S}^{3}(r_{S}-r_{B})},
\end{equation}
which determines the Page time as
\begin{equation}
t_{\text{Page}}\simeq\dfrac{48\pi^{2}r_{S}^{3}}{c\kappa_{4}^{2}}=\dfrac{3S_{\text{BH}}}{\pi cT_{\text{H}}}.
\end{equation}
In this way, the Page time is proportional to the ratio of the black string entropy and the Hawking temperature, which are the thermodynamic quantities of the black string. This result shows a universal behavior of the Page time, which has been found in various black hole geometries \cite{27, Ageev,LiWang2021,KimNam2021,Karananas2021,YuGe2021,Ahn2021,13,HNam2022}. Indeed, we can apply a simple argument given by D. Page \cite{Page1993a,Page1993b}: for a total system that consists of a sufficiently small subsystem, the entanglement entropy can be approximately determined by the thermal entropy of that subsystem. With the black string having a large entropy, it needs to take more time in order for a large amount of the radiation to enter into the cutoff surface in such a way that the black hole is a subsystem at late times and according to the argument of D. Page the entanglement entropy would be approximated by the black hole entropy. This situation is also similar to the black hole of low temperature, which requires more time for the evaporation to become a subsystem that is small enough compared to the total system.

Because the island is in the entanglement wedge of radiation, the signal falling into the black hole would be decoded from the outgoing radiation when this signal enters the island. We can calculate the scrambling time, which is defined as the minimum duration to retrieve the information of the signal falling into the black hole according to the Hayden-Preskill protocol \cite{Hayden2007}. Suppose that one sends a signal at the time $t=0$ from the cutoff surface at $ r=b $ to the island boundary $r=a$, the minimal time that it takes is approximately the scrambling time and is givens as follows

\begin{equation}
\begin{split}
t_{\text{scr}}&=r_{*}(b)-r_{*}(a)\\
&=-a\sqrt{1-\dfrac{r_{B}}{a}}+b\sqrt{1-\dfrac{r_{B}}{b}}+\left(\dfrac{r_{B}}{2}+r_{S}\right) \log\left[-1+\dfrac{2b}{r_{B}}\left(1+\sqrt{1-\dfrac{r_{B}}{b}} \right) \right]\\
&\quad -\left(\dfrac{r_{B}}{2}+r_{S}\right)\log\left[-1+\dfrac{2a}{r_{B}}\left(1+\sqrt{1-\dfrac{r_{B}}{a}} \right) \right]   \\
&\quad +\dfrac{r_{S}^{3/2}}{\sqrt{r_{S}-r_{B}}}\log \left[ \dfrac{2b}{b-r_{S}}\sqrt{1-\dfrac{r_{B}}{b}}+\dfrac{b(r_{B}-2r_{S})+r_{B}r_{S}}{(r_{S}-b)\sqrt{r_{S}(r_{S}-r_{B})}}\right] \\
&\quad -\dfrac{r_{S}^{3/2}}{\sqrt{r_{S}-r_{B}}}\log \left[ \dfrac{2a}{a-r_{S}}\sqrt{1-\dfrac{r_{B}}{a}}+\dfrac{a(r_{B}-2r_{S})+r_{B}r_{S}}{(r_{S}-a)\sqrt{r_{S}(r_{S}-r_{B})}}\right] \\
&\simeq\dfrac{1}{\kappa} \log \dfrac{\sqrt{r_{S}^{3}(r_{S}-r_{B})}}{\kappa_{4}^{2}}\\
&\simeq \dfrac{1}{2\pi T_{\text{H}}}\log S_{\text{BH}}.
\end{split}
\end{equation}
The black string entropy $S_{\text{BH}}$ can be expressed in terms of the number of microstates $N$ as $S_{\text{BH}}=e^N$. With respect to the macroscopic scale black string, the number of microstates is very large. Therefore, due to the scrambling time proportional to the logarithm of the black hole entropy, the scrambling time is much smaller than the Page time. 

\section{\label{CHD}The case of higher dimensions}
\label{sec:4}

In this section, we calculate the entanglement entropy of the Hawking radiation, which is emitted from the higher-dimensional non-singular black strings relying on the island formula. By generalizing the five-dimensional regular black string solution to higher dimensions, one can find 
\begin{equation}
ds_{D+1}^{2}=-f_{S}(r)dt^{2}+f_{B}(r)dy^{2}+\dfrac{dr^{2}}{f_{S}(r)f_{B}(r)}+r^{2}d\Omega_{D-2}^{2},
\end{equation}
where
\begin{equation}
f_{B}(r)=1-\left( \dfrac{r_{B}}{r}\right) ^{D-3},\quad f_{S}(r)=1-\left( \dfrac{r_{S}}{r}\right) ^{D-3}.
\end{equation}
The temperature and the entropy of the higher-dimensional black strings read
\begin{eqnarray}
 T_\text{H}&=&\frac{\kappa}{2\pi}=\dfrac{D-3}{4\pi r_{S}}\sqrt{1-\left( \dfrac{r_{B}}{r_{S}}\right)^{D-3} },\nonumber\\
S_{\text{BH}}&=&\dfrac{4\pi^{\dfrac{D+1}{2}}}{\Gamma\left( \dfrac{D-1}{2}\right)\kappa_{D}^{2} }\sqrt{r_{S}^{D-1}\left( r_{S}^{D-3}-r_{B}^{D-3}\right)}.
\end{eqnarray}

First, we need to determine the tortoise coordinate which is given as follows
\begin{equation}
\begin{split}
r_{*}(r)&=\int \dfrac{dr}{\left[1-\left( \dfrac{r_{S}}{r}\right)^{D-3}  \right]\sqrt{1-\left( \dfrac{r_{B}}{r}\right)^{D-3} } }\\
&\simeq \int \dfrac{dr}{\left[1-\left( \dfrac{r_{S}}{r}\right)^{D-3}  \right]\sqrt{1-\left( \dfrac{r_{S}}{r}\right)^{D-3} \delta} }\\
&\simeq {_2F_1}\left( 1,\dfrac{1}{3-D},\dfrac{D-4}{D-3},\left( \dfrac{r_{S}}{r}\right)^{D-3} \right)r\\
&\quad-{_2F_1}\left( 1,\dfrac{D-4}{D-3},1+\dfrac{D-4}{D-3},\left( \dfrac{r_{S}}{r}\right)^{D-3} \right)\dfrac{r}{2(D-4)}\left(\dfrac{r_{S}}{r} \right)^{D-3}\delta,\label{HD-tor-coor}
\end{split}
\end{equation}
where $\delta\equiv r_{B}/r_{S} $ and ${_2F_1}(a,b;c;z)$ is the hypergeometric function. Note that, the integral at the first line of Eq. (\ref{HD-tor-coor}) in the general case of $D>4$ is difficult to be exactly obtained. But, in the situation $ r_{B}\ll r_{S} $ corresponding to $\delta\ll1$
which means that the bubble lies deep into the black string, we can find an analytical expression for the tortoise coordinate which is given in terms of the hypergeometric functions.

With the configuration without the island, it is easy to calculate the entanglement entropy which is given in late times as follows
\begin{equation}
S_{R}\simeq \dfrac{c(D-3)}{6r_{S}}\sqrt{1-\left( \dfrac{r_{B}}{r_{S}}\right)^{D-3} }t_{b}= \dfrac{c}{3}\kappa t_{b}.
\end{equation}
This expression implies that the entanglement entropy without the contribution of the island also evolves linearly in time in analogy to the five-dimensional case. Consequently, it results in the information loss problem in the evaporation of the higher-dimensional non-singular black string.

Next, we study how the island configuration contributes to the entanglement entropy and makes the black string evaporation consistent with the unitarity of quantum mechanics. At early times, the entanglement entropy of the black string with the radiation is small and thus the island has to be deeply hidden inside the horizon if it exists. The generalized entropy can be approximately calculated as
\begin{equation}
S_{\text{gen}}\simeq\dfrac{8\pi^{\dfrac{D+1}{2}}}{\Gamma\left( \dfrac{D-1}{2}\right) \kappa_{D}^{2}}\sqrt{a^{D-1}(a^{D-3}-r_{B}^{D-3})}+\dfrac{c}{6}\log\left[ \left(\dfrac{r_{S}}{a} \right)^{D-3}-1 \right] .
\end{equation}
By extremizing the entropy over $ a $, we find the position of the island as
\begin{equation}
a^{D-2}\sim c\kappa_{D}^{2},
\end{equation}
which is in the order of the quantum gravity scale. This means that the island does not emerge at early times.

At late times, we expect the appearance of the island when more and more Hawking radiation is emitted from the black string in order to lead to a transition of the entanglement entropy from linear growth due to the absence of the island to the constant behavior. With the approximation of late times given by Eq. (\ref{latet-appr}), we can calculate the generalized entropy as
\begin{equation}
\begin{split}
S_{\text{gen}}\simeq &\dfrac{8\pi^{\dfrac{D+1}{2}}}{\Gamma\left( \dfrac{D-1}{2}\right)\kappa_{D}^{2} }\sqrt{a^{D-1}\left( a^{D-3}-r_{B}^{D-3}\right)}\\
&+\dfrac{c}{6}\left[ W^{2}(a)W^{2}(b)\right]+\dfrac{c}{3}\log \left[ 1-2e^{-\kappa(r_{*}(b)-r_{*}(a))}\right]\\
\simeq & \dfrac{8\pi^{\dfrac{D+1}{2}}}{\Gamma\left( \dfrac{D-1}{2}\right)\kappa_{D}^{2} }\sqrt{a^{D-1}\left( a^{D-3}-r_{B}^{D-3}\right)}\\
&+\dfrac{c}{6}\log \left[ f_{S}(a)f_{S}(b) \right]-\dfrac{c}{3} \kappa r_{*}(a)+\dfrac{c}{3} \kappa r_{*}(b)-\dfrac{2c}{3}e^{-\kappa(r_{*}(b)-r_{*}(a))}.
\end{split}
\end{equation}
In order to find the island location which is slightly outside the black hole, we write the radius of the island as $ a=r_{S}+\epsilon $ with $ \epsilon\ll1$. Then, by extremizing the generalized entropy, we derive an equation that determines the position of the island's boundary at the first-order approximation as
\begin{equation}
\begin{split}
&\dfrac{8\pi^{\dfrac{D+1}{2}}}{\Gamma\left( \dfrac{D-1}{2}\right)}\dfrac{ r_{S}^{D-3}}{c\kappa_{D}^{2}}\left(D-2-\dfrac{X}{2} \right) -\dfrac{(D-2)}{6r_{S}}+\dfrac{(D-3)X}{12r_{S}}\\
&-\dfrac{2L}{3}\dfrac{D-3}{2r_{S}}\sqrt{\dfrac{D-3}{r_{S}\epsilon}}\left[ \dfrac{r_{S}}{(D-3)}+\dfrac{D-2}{2(D-3)}\epsilon-\dfrac{(D-4)X}{4(D-3)}\epsilon\right]=0 ,\label{equ-eps-HD}\\
\end{split}
\end{equation}
where $X$ and $L$ are defined as
\begin{eqnarray}
X&\equiv&\left(\dfrac{r_{B}}{r_{S}} \right) ^{D-3},\\
 L&\equiv&\exp\left\{\kappa\left(\dfrac{r_{S}}{D-3}\left[\gamma+P\left(0,\dfrac{1}{3-D}\right)\right]-r_{*}(b)\right)\right\},
\end{eqnarray}
where $ \gamma\simeq 0.577 $ is Euler's constant and $ P\left(0,(3-D)^{-1}\right)\equiv\Gamma'\left((3-D)^{-1}\right)/ \Gamma\left((3-D)^{-1}\right)$ is the polygamma function of order 0 with $\Gamma'(z)$ denoted the first derivative of the Gamma function $\Gamma(z)$. With respect to the terms independent on $\epsilon$, we observe that due to $ r_{S}^{D-2}/c\kappa_{D}^{2}\gg 1 $, the second and third terms are much smaller than the first term in the left-hand side of Eq. (\ref{equ-eps-HD}) and they hence can be ignored. For the terms relating to $\epsilon$, the final two terms in the bracket can be eliminated for the same reason. As a result,  we find an approximation equation determining the position of the island's boundary as follows
\begin{equation}
\dfrac{8\pi^{\dfrac{D+1}{2}}}{\Gamma\left( \dfrac{D-1}{2}\right)}\dfrac{ r_{S}^{D-3}}{c\kappa_{D}^{2}}\left(D-2-\dfrac{X}{2} \right)-\dfrac{2L}{3}\dfrac{D-3}{2r_{S}}\sqrt{\dfrac{D-3}{r_{S}}\epsilon} \dfrac{r_{S}}{(D-3)\epsilon}=0.
\end{equation}
Solving this equation leads to the location of the island's boundary as
\begin{equation}
a=r_{S}+\dfrac{c^{2}\kappa_{D}^{4}}{r_{S}^{2D-4}}\dfrac{(D-3)L^{2}\Gamma\left( \dfrac{D-1}{2}\right)^{2} }{576(D-2)^{2}\pi^{D+1}}\left[1+\dfrac{X}{D-2} \right] r_{S}.
\end{equation}
Because of $c^{2}\kappa_{D}^{4}/r_{S}^{2D-4}\ll1$, indeed the island locates slightly outside the event horizon. The dependence of the island position on the radius of the bubble $ r_{B} $ is manifested via the quantity $ X $. The island goes further from the black hole when the bubble's radius increases. This is compatible with the behavior of the five-dimensional case in the regime of $r_B/r_S\ll1$.

Substituting the island position into the generalized entropy, we obtain the entanglement entropy of the radiation as follows
\begin{equation}
\begin{split}
S_{\text{EE}}&\simeq \dfrac{8\pi^{\dfrac{D+1}{2}}}{\Gamma\left( \dfrac{D-1}{2}\right)\kappa_{D}^{2} }\sqrt{r_{S}^{D-1}\left( r_{S}^{D-3}-r_{B}^{D-3}\right)}\\
&\quad+\dfrac{c}{6}\log\left[ \dfrac{(D-3)L^{2}\Gamma\left( \dfrac{D-1}{2}\right)^{2} }{576(D-2)^{2}\pi^{D+1}}\left(1+\dfrac{X}{D-2} \right)\right] + \mathcal{O}\left(\dfrac{c^{2}\kappa_{D}^{4}}{r_{S}^{2D-4}} \right) .
\end{split}
\end{equation}
The first term is twice the Bekenstein-Hawking entropy which comes from the contribution of the two-sided island area and is a dominant part of the entanglement entropy. The second term is the logarithmic correction which comes from the quantum nature of the radiation. The higher-order correction terms are very small due to the power factors of the tiny ratio $ c\kappa_{D}^{2}/r_{S}^{D-2} $. Therefore, at the leading order the entanglement entropy at late times is twice the Bekenstein-Hawking entropy which is a finite constant due to the appearance of the island. This result manifests that the entanglement entropy is bounded by the Bekenstein-Hawking entropy instead of growing linearly in time. This means that the information of the higher-dimensional non-singular black string is preserved during the evaporation. 

\textbf{Page time and scrambling time}. Although the entanglement entropy of the Hawking radiation that is emitted from the higher-dimensional non-singular black string grows linearly in time at the early stage, with the appearance of the island at late times the entanglement entropy reaches the maximum. The moment that determines the transition between these behaviors is given by the Page time as
\begin{equation}
t_{\text{Page}}\simeq \dfrac{3S_{\text{BH}}}{\pi cT_{\text{H}}},
\end{equation}
which depends on the ratio of the entropy to the temperature of the higher-dimensional non-singular black string. This dependence is universal at the leading order independent of the spacetime dimension. It is interesting that although both $S_{\text{BH}}$ and $T_\text{H}$ are dependent on the bubble radius $r_B$, at the leading order the Page time is not affected by the change of the bubble radius. The presence of the island implies that the information entering into the island is retrievable from the Hawking radiation. In this situation, it is easy to calculate the scrambling time corresponding to the information recovery time of the signal when it falls into the black string from the cutoff surface at $t=0$ and reaches the island boundary as
\begin{equation}
\begin{split}
t_{\text{scr}}&=r_{*}(b)-r_{*}(a)\\
&\simeq {_2F_1}\left( 1,\dfrac{1}{3-D},\dfrac{D-4}{D-3},\left( \dfrac{r_{S}}{b}\right)^{D-3} \right)b - {_2F_1}\left( 1,\dfrac{1}{3-D},\dfrac{D-4}{D-3},\left( \dfrac{r_{S}}{a}\right)^{D-3} \right)a\\
&\simeq \dfrac{1}{2\pi T_{\text{H}}}\log S_{\text{BH}}.
\end{split}
\end{equation}
The scrambling time is proportional to the ratio of the logarithm of the Bekenstein-Hawking entropy to the Hawking temperature. This dependence is universal at the leading order approximation. In order to see the effect of the bubble on the scrambling time, let us expand $t_{\text{scr}}$ in terms of $r_B/r_S$ as
\begin{eqnarray}
 \frac{D-3}{r_S}t_{\text{scr}}\simeq2\log\left[\frac{4\pi^{\frac{D+1}{2}}r^{D-2}_S}{\Gamma\left( \dfrac{D-1}{2}\right)\kappa_{D}^{2}}\right]+\left\{1-\log\left[\frac{4\pi^{\frac{D+1}{2}}r^{D-2}_S}{\Gamma\left( \dfrac{D-1}{2}\right)\kappa_{D}^{2}}\right]\right\}\left(\frac{r_B}{r_S}\right)^{D-3}.
\end{eqnarray}
We see that due to $r^{D-2}_S/\kappa_{D}^{2}\gg1$ the second term is always negative. This means that in the region $r_B/r_S\ll1$ the presence of the bubble behind the event horizon would make the information recovery of the signal falling into the black string faster.  

\section{\label{conclu} Conclusion}
The unphysical curvature singularity, the nature of microstates associated with the Bekenstein-Hawking entropy, and the information loss paradox are three important problems of black hole physics. It is expected that these problems would be solved in a consistent theory of quantum gravity that needs to describe the quantum fluctuations of spacetime that become important in the region close to the center of black holes. Currently, there is still no complete ultraviolet theory of quantum gravity. But, it is reasonable to expect that the traces of quantum gravity in the low-energy regime which provide indications toward the resolution of the critical puzzles. In this work, we point out that the compactified extra dimensions and the entanglement islands may be important clues to seeking a consistent theory of quantum gravity because they can tell how all deepest puzzles of black hole physics are solved in quantum gravity.

In the five-dimensional Einstein-Maxwell theory with the extra dimension compactified on a circle $S^1$, one can find two vacuum solutions that are a four-dimensional Schwarzschild black hole times $S^1$ and a static bubble of nothing (or smooth massless solution). By imposing these vacuum solutions to be symmetric under the double Wick rotation between the usual time dimension and the compactified extra dimension, the non-singular black string and the smooth bubble solution (topological star) have been found by turning on the appropriate magnetic fluxes \cite{8}. The usual curvature singularity is naturally removed due to the presence of a bubble hidden behind the event horizon which ends spacetime. Additionally, the smooth bubble solution, black string, and RN black hole can live together in a regime of mass and charge, which implies that the smooth bubble geometries can be realized as microstates leading to the Bekenstein-Hawking entropy of the regular black string.

The gravitational Euclidean path integral (which is one of the main methods to explore quantum gravity) with the replica wormholes as new saddle points leads to the emergence of the island configuration. Taking into account the contribution of the island to calculate the entanglement entropy of the radiation that is emitted from the regular black string, we show that the entanglement entropy follows the Page curve consistent with the unitarity of quantum mechanics during the evaporation process of the regular black string. At early times of the evaporation, the entanglement entropy increases linearly with time which implies the information loss paradox. However, when the radiation entering the cutoff surface becomes more and more at late times, the island emerges and locates nearly outside the event horizon. As a result, the entanglement entropy reaches a saturation value that is approximate twice the Bekenstein-Hawking entropy of the regular black string. We calculate the Page time that determines the transition between the linearly growing behavior and the convergent behavior of the entanglement entropy. The leading order term of the Page time is universal and it is not affected by the change in the bubble radius. Because the signal that falls into the black hole is decoded from the outgoing radiation when it enters the island, we compute the scrambling time according to the Hayden-Preskill protocol \cite{Hayden2007}. We find that the scrambling time at the leading order is proportional to the ratio of the logarithm of the Bekenstein-Hawking entropy to the Hawking temperature. Because of this relationship, unlike the Page time, the scrambling time is dependent on the presence of the bubble and hence it would be affected due to the change in the bubble radius.

\end{document}